\documentclass[12pt]{article}
\usepackage{latexsym}
\usepackage{amsmath,,calrsfs}
\usepackage{amsfonts}
\usepackage{amssymb}
\usepackage{amscd}
\usepackage{bbm}
\usepackage{fancybox}
\usepackage{cite}
\usepackage{amsmath,amsfonts,amsbsy}
\usepackage{pstricks,pst-node}
\usepackage[small,bf,hang]{caption2}
\usepackage{graphicx}
\usepackage{epsfig}
\usepackage{psfrag}
\usepackage{comment}

\usepackage{float}

\psset{unit=1.3cm,linewidth=.5pt,radius=.2}  

\usepackage{multirow}                     
\usepackage{float}                          
\usepackage{lscape}                         
\usepackage{bm}

\newcommand{\beq}{\begin{equation}}
\newcommand{\eeq}{\end{equation}}
\newcommand{\bi}{\begin{itemize}}
\newcommand{\ei}{\end{itemize}}
\newcommand{\bt}{\begin{tabular}}
\newcommand{\et}{\end{tabular}}
\newcommand{\bc}{\begin{center}}
\newcommand{\ec}{\end{center}}

\newcommand{\be}{\begin{equation}}
\newcommand{\ee}{\end{equation}}
\newcommand{\bea}{\begin{eqnarray}}
\newcommand{\eea}{\end{eqnarray}}
\newcommand{\ba}{\begin{array}}
\newcommand{\ea}{\end{array}}

\def\bbox{{\,\lower0.9pt\vbox{\hrule \hbox{\vrule height 0.2 cm
\hskip 0.2 cm \vrule height 0.2 cm}\hrule}\,}}
\newcommand{\dsl}{\pa \kern-0.5em /}

\makeatletter \@addtoreset{equation}{section} \makeatother

\def\slashchar#1{\setbox0=\hbox{$#1$}           
   \dimen0=\wd0                                 
   \setbox1=\hbox{/} \dimen1=\wd1               
   \ifdim\dimen0>\dimen1                        
      \rlap{\hbox to \dimen0{\hfil/\hfil}}      
      #1                                        
   \else                                        
      \rlap{\hbox to \dimen1{\hfil$#1$\hfil}}   
      /                                         
   \fi}

\begin{document}

\begin{titlepage}
\begin{center}

\hfill UG-11-23, DAMTP-2011-58

\vskip 1.5cm

{\Large \bf  A spin-4 analog of 3D massive gravity
}

\vskip 1cm

{\bf Eric A.~Bergshoeff\,${}^1$\,, Marija Kovacevic\,${}^1$\,,
Jan Rosseel\,${}^1$\,, \\
Paul K.~Townsend\,${}^2$ and Yihao Yin\,${}^1$} \\

\vskip 25pt

{\em $^1$ \hskip -.1truecm
\em Centre for Theoretical Physics,
University of Groningen, \\ Nijenborgh 4, 9747 AG Groningen, The
Netherlands \vskip 5pt }

{email: {\tt E.A.Bergshoeff@rug.nl, maikovacevic@gmail.com,
j.rosseel@rug.nl, y.yin@rug.nl}} \\

\vskip .4truecm

{\em $^2$ \hskip -.1truecm
\em  Department of Applied Mathematics and Theoretical Physics,\\ Centre for Mathematical Sciences, University of Cambridge,\\
Wilberforce Road, Cambridge, CB3 0WA, U.K.\vskip 5pt }

{email: {\tt P.K.Townsend@damtp.cam.ac.uk}} \\

\end{center}

\vskip 0.5cm

\begin{center} {\bf ABSTRACT}\\[3ex]
\end{center}

A $6$th-order, but ghost-free, gauge-invariant action is found for a $4$th-rank symmetric tensor potential  in a three-dimensional (3D)  Minkowski spacetime.  It propagates  two massive modes of spin $4$ that are interchanged by  parity, and is thus a spin-$4$ analog of linearized   ``new massive gravity''.  Also found are
ghost-free spin-$4$ analogs of  linearized ``topologically massive gravity'' and ``new topologically massive gravity'', of 5th- and 8th-order respectively.


\end{titlepage}

\newpage
\setcounter{page}{1} \tableofcontents

\newpage

\section{Introduction}\label{sec:intro}

There is a well-developed theory of relativistic free-field spin-$s$ gauge theories in a 4-dimensional Minkowski spacetime (4D), based on symmetric rank-$s$ gauge potentials. The topic was initiated by Fronsdal \cite{Fronsdal:1978rb} and its geometric formulation was provided by de Wit and Freedman  \cite{deWit:1979pe}. We refer the reader  to \cite{Bouatta:2004kk} for a more recent review.
The $s\le2$ cases are standard; in particular,  the $s=2$ field equation is the linearized Einstein equation for a metric perturbation. This provides a model for  integer ``higher-spin'' ($s>2$) where the gauge-invariant two-derivative field-strength is an analog of the linearized Riemann tensor.  A feature of these higher-spin gauge theories of relevance here is that the gauge transformation parameter, a symmetric tensor of rank $s-1$, is  constrained to be trace-free. If this constraint on the parameter were to be relaxed then any  gauge-invariant equation would be of higher than second order, and this would normally imply the propagation of ghost modes, i.e. modes of negative energy.

The situation for 3-dimensional Minkowski spacetime (3D) is different, in many respects.  One is that the standard ``higher-spin''  gauge field equations  do not actually propagate any modes in 3D. One may take advantage of this simplification, and the fact that 3D gravity can be recast  as a Chern-Simons (CS) theory \cite{Achucarro:1987vz,Witten:1988hc},  to construct  CS models for higher-spin fields interacting with 3D gravity in an anti de Sitter (adS) background. The original model of this type  \cite{Blencowe:1988gj} is analogous to
Vasiliev's 4D theory of {\it all} integer higher-spins interacting in  an adS background  \cite{Vasiliev:1990en}. However, in 3D one can consider a ``truncated'' version describing only a finite number of higher-spin fields coupled to gravity  \cite{Henneaux:2010xg,Campoleoni:2010zq}.  Such models have recently yielded interesting insights  \cite{Castro:2010ce,Gutperle:2011kf,Ammon:2011nk}
although the absence of propagating modes  may limit their impact.

Propagating modes arise in 3D when higher-derivative terms are included in the action. The best known case is
``topologically massive gravity'' (TMG) which involves the inclusion of a $3$rd-order Lorentz-Chern-Simons term \cite{Deser:1981wh}. This is a parity-violating gravity model that propagates a single massive spin-$2$ mode, thereby illustrating another special feature of 3D:  gauge-invariance is consistent with non-zero mass. TMG is ghost-free, despite the higher-derivative nature of the field equations, because one may choose the overall sign of the action to ensure that the one propagated mode has positive energy.  Rather more surprising is  the fact that there exists a parity-preserving unitary model with curvature-squared terms, and hence $4$th-order equations, that is ghost-free and propagates two massive spin-$2$ modes, which are exchanged by parity;  this is  ``new massive gravity'' (NMG) \cite{Bergshoeff:2009hq}.   It is notable that the problems normally associated with non-linearities in higher-derivative theories  \cite{Boulware:1973my} are absent in NMG  \cite{deRham:2011ca}.

These facets of gauge-field dynamics in 3D are, by now, well known. Less well known, because it is peculiar to ``higher spin'' ($s>2$)  is yet another unusual feature: the trace-free constraint on the gauge parameter may be relaxed, resulting in what we shall call an ``unconstrained''  higher-spin gauge invariance. As stated above, this implies higher-order  field equations, but this need not imply a violation of unitarity in 3D. The spin-$3$ case was discussed in  \cite{Bergshoeff:2009tb}. Two distinct parity-violating ghost-free spin-$3$ models with unconstrained gauge invariance were found there. One is a natural spin-$3$ analog of TMG and, as for TMG, the absence of ghosts is essentially a consequence of the fact that  only one mode is propagated. Nevertheless,  the unconstrained nature of the gauge invariance is crucial; a previous attempt to construct a spin-$3$ analog of TMG with a tracefree gauge parameter led to a model propagating an additional spin-$1$ ghost \cite{Damour:1987vm}. We should point out here that
another more recent model has also been called a  ``Spin-$3$ TMG''  \cite{Chen:2011vp}.

The above considerations motivate the investigation of higher-spin  gauge theories in 3D with unconstrained gauge invariance.
A systematic procedure for the construction of  such  theories  was proposed in \cite{Bergshoeff:2009tb}. Starting from the 3D version of the standard massive Fierz-Pauli (FP)  equations for a rank-$s$ symmetric tensor field,   which we denote by $G$, one can solve the subsidiary differential constraint on this field to obtain an expression\footnote{Here we use a normalization different from that used in  \cite{Bergshoeff:2009tb}.} for it  in terms of a rank-$s$ symmetric tensor gauge potential $h$:
\be
G_{\mu_1\dots \mu_s}  = \varepsilon_{\mu_1}{}^{\tau_1\nu_1} \cdots \varepsilon_{\mu_s}{}^{\tau_s\nu_s} \partial_{\tau_1} \cdots\partial_{\tau_s} \ h_{\nu_1\cdots \nu_s}\, .
\ee
We now view $G$ as the field strength for $h$; it is invariant under the gauge transformation
\be
\delta_\xi h_{\mu_1\cdots \mu_s} = \partial_{(\mu_1}\xi_{\mu_2\cdots\mu_s)}\, ,
\ee
where the infinitesimal symmetric tensor parameter of rank-$(s-1)$ is {\it unconstrained}. The FP equations become
\be\label{FPequiv}
\left(\square -m^2\right) G_{\mu_1 \cdots \mu_s} =0 \, , \qquad \eta^{\mu\nu} G_{\mu\nu\rho_1\cdots \rho_{s-2}}=0\, .
\ee
The dynamical equation is now of order $s+2$, and hence ``higher derivative'' for $s>0$.  What was the algebraic constraint  is now a differential constraint  of order $s$, and what was  the differential constraint is now the Bianchi identity
\be\label{Bianchi}
\partial^\nu G_{\nu\mu_1\cdots \mu_{s-1}} \equiv 0\, .
\ee
This procedure can equally be applied to the  parity-violating ``square-root FP'' ($\sqrt{\mathrm{FP}}$) spin-$s$ equations that propagate one mode rather than two. In this case, taking the mass to be $\mu$,  one gets the  topologically-massive spin-$s$ equations
\be\label{sqrtFP}
\varepsilon_{\mu_1}{}^{\tau\lambda} \partial_\tau G_{\lambda\mu_2 \cdots \mu_s} =  \mu G_{\mu_1 \mu_2\cdots \mu_s} \, , \qquad
\qquad \eta^{\mu\nu} G_{\mu\nu\rho_1\cdots \rho_{s-2}}=0\, .
\ee
For $s=2$ these are the equations of linearized TMG and for $s=3$ they are the equations of the spin-$3$ analog of TMG mentioned above.

Given the equivalence of the unconstrained gauge theory formulation of spin-$s$ field equations with the standard  FP and $\sqrt{\mathrm{FP}}$ equations, one may  ask what is to be gained by a gauge theory  formulation: what advantage does it  have over the original FP formulation?  In the $s=2$ case the answer is that it allows the introduction of local interactions, through a gauge principle, that would otherwise be impossible: linearized NMG is the linearization of the non-linear NMG, which is {\it not} equivalent to any non-linear modification of the FP theory of massive spin $2$ (and the same is true of TMG).  One may hope for something similar in the higher-spin case, although we expect this to be much less straightforward. It may be necessary to consider  all even spins, or an adS background, as in Vasiliev's 4D theory. There is also a potential link to new 3D string theories \cite{Mezincescu:2010yp}.

The linear gauge-theory equations (\ref{FPequiv}) propagate, by construction, two spin-$s$ modes that are interchanged by parity, but the construction only guarantees an {\it on-shell}  equivalence with FP theory. There is no guarantee that  both spin-$s$ modes  are physical,  rather than ghosts; this depends on the signs of the kinetic energy terms in an (off-shell) action\footnote{As already mentioned, this is not an issue for  ``topologically-massive'' theories because they propagate only one mode and the sign of the action may be chosen such that this one mode is physical.}. The action  that yields the $s=1$ case of (\ref{FPequiv}) has been studied previously as  ``extended topologically massive electrodynamics''  (ETME) \cite{Deser:1999pa}, and one of the two spin-$1$ modes turns out to be a ghost, so the on-shell equivalence to FP does not extend to an off-shell equivalence for $s=1$.  In contrast, the $4$th-order spin-$2$ equations are those of linearized  NMG, for which both spin-$2$ modes  are physical.  Moving on to $s=3$, the construction of an action shows that one of the spin-$3$ modes is a ghost \cite{Bergshoeff:2009tb}, exactly as for spin $1$.

No attempt to construct actions for $s\ge4$ was made in \cite{Bergshoeff:2009tb} because this requires additional ``auxiliary''  fields. Here we construct, by finding the required auxiliary fields,  actions that can be described as spin-$4$ analogs of TMG and NMG. In the latter case, the absence of ghosts is a non-trivial issue, which we settle using the method introduced  by Deser for NMG \cite{Deser:2009hb} and further developed in  \cite{Andringa:2009yc}. This result is consistent with a conjecture of  \cite{Bergshoeff:2009tb}, on which we elaborate at the end of this paper,  that a spin-$s$ analog of NMG (i.e. a  ghost-free parity-invariant action of order  $s+2$ in derivatives propagating two spin-$s$ modes)  exists  only for {\it even} $s$.

Although it might seem remarkable that  a 6th-order action for spin-$4$ can be ghost-free, it is possible to
construct (linear) higher-order ghost-free spin-$4$ models by enlarging the gauge invariance to include
a spin-$4$ analog of linearized spin-$2$ conformal invariance, with a symmetric 2nd-rank tensor parameter.
In fact, spin-$s$ gauge field equations  of this type can be found by solving simultaneously both the differential subsidiary condition of 
the FP or $\sqrt{\mathrm{FP}}$ theory and its algebraic trace-free condition \cite{Bergshoeff:2009tb}, and these equations may be integrated to an action without the need for auxiliary fields.  For example, the  spin-$s$ $\sqrt{\mathrm{FP}}$ equations become
the equations that follow by variation of the symmetric rank-$s$ tensor $h$ in the action 
\be\label{confsact}
S[h] = \frac{1}{2} \int d^3 x \left\{ h^{\mu_1\dots\mu_s} C_{\mu_1\dots\mu_s} 
+ \frac{1}{\mu} \varepsilon_\mu{}^{\alpha\beta} 
h^{\mu\nu_1\dots\nu_{s-1}} \partial_\alpha C_{\beta\nu_1\dots \nu_{s-1}}
\right\}\, , 
\ee
where $C$ is the spin-$s$ Cotton-type tensor for $h$ \cite{Pope:1989vj}, defined  (up to a factor) as the rank-$s$ symmetric tensor of order $(2s-1)$ in derivatives that is invariant under the spin-$s$ generalization of a linearized conformal gauge transformation:
\be
\delta_\Lambda h_{\mu_1\mu_2\mu_3 \dots \mu_s} = \eta_{(\mu_1\mu_2} \Lambda_{\mu_3\dots \mu_s)}\, . 
\ee
A convenient expression for the Cotton-type tensor is
\be
C_{\mu_1\dots\mu_s} = \varepsilon_{\mu_1\dots\mu_s} = \varepsilon_{(\mu_1}{}^{\nu_1\rho_1} \cdots \varepsilon_{\mu_{s-1}}{}^{\nu_{s-1}\rho_{s-1}}\partial_{|\nu_1} \cdots \partial_{\nu_{s-1}} S_{\rho_1\dots\rho_{s-1}| \mu_s)}\, , 
\ee
where the rank-$s$ symmetric tensor $S$, of order $s$ in derivatives,  is a spin-$s$ generalization of the linearized 3D Schouten tensor with the conformal-type transformation 
\be
\delta_\Lambda S_{\mu_1\mu_2\mu_3\dots \mu_s} = \partial_{(\mu_1}\partial_{\mu_2} \Omega_{\mu_3\dots\mu_s)}\, , 
\ee
where $\Omega$ is a rank-$(s-2)$ tensor operator, of order $(s-2)$ in derivatives, acting on the rank-$(s-2)$ tensor parameter 
$\Lambda$. 

Applied to the $\sqrt{\mathrm{FP}}$ spin-$2$ model, this construction yields the linearized 4th-order  ``new topologically massive gravity'' theory of \cite{Andringa:2009yc}, found and analysed independently in \cite{Dalmazi:2009pm}.  It was also used in \cite{Bergshoeff:2009tb} to find a 6th-order  ghost-free action for a single spin-$3$ mode. Here we present details of  the $s=4$ case. In particular, we verify that the $8$th-order spin-$4$ action of the type (\ref{confsact}) propagates a single mode and we show that it  is ghost-free. We also apply the construction to the spin-$4$ FP theory, obtaining a $9$th parity-preserving action but in this case one of the two spin-$4$ modes  is a ghost, as for the analogous lower-spin cases considered in \cite{Bergshoeff:2009tb}.

\section{Spin-$4$ equations}

Setting $s=4$ in (\ref{sqrtFP}) we obtain the ``topologically-massive''  equations for a single spin-$4$ mode of mass $\mu$, described by a
$4$th-rank symmetric gauge potential $h$:
\begin{equation}\label{sqrtFP4}
\varepsilon_\mu{}^{\tau\lambda} \partial_\tau G_{\lambda\nu\rho\sigma} = \mu G_{\mu\nu\rho\sigma} \, , \qquad
G^{\text{tr}}_{\mu \nu }(h)=0\, ,
\end{equation}
where we have defined $G^{\text{tr}}_{\mu \nu }(h) = \eta^{\rho\sigma}G_{\mu\nu\rho\sigma}(h)$.
Similarly, setting $s=4$ in  (\ref{FPequiv}) we obtain the parity-preserving field equations for a pair of spin-$4$ modes of mass $m$, exchanged by parity\footnote{According to one definition, the spin of a particle in 3D may have either sign, which is flipped by parity. We
call this the ``relativistic helicity'' and define spin  to be its absolute value. According to this definition,  both modes of a parity doublet have the same spin.}
\be\label{FPequiv4}
\left(\square -m^2\right) G_{\mu\nu\rho\sigma}(h) =0\, , \qquad G^{\text{tr}}_{\mu \nu }(h)=0\, .
\ee
In either case,
\be\label{fieldstrength}
G_{\mu\nu\rho\sigma}(h) = \varepsilon_{\mu}{}^{\tau\alpha} \varepsilon_\nu{}^{\eta\beta} \varepsilon_\rho{}^{\xi\gamma}
\varepsilon_\sigma{}^{\zeta\delta} \partial_\tau\partial_\eta\partial_\xi\partial_\zeta \, h_{\alpha\beta\gamma\delta}
 \, , \qquad G^{\text{tr}}_{\mu \nu }(h) = \eta^{\rho\sigma}G_{\mu\nu\rho\sigma}(h)\, .
\ee
We use a ``mostly plus'' metric convention and define the isotropic rank-$3$ antisymmetric  tensor $\varepsilon$ such that\footnote{This sign is opposite to that used in  \cite{Andringa:2009yc} and \cite{Bergshoeff:2009hq}.}
\begin{equation}
\varepsilon^{012} =-1\, .
\end{equation}
The tensor $G(h)$ is  invariant under the gauge  transformation
\be
\delta h_{\mu\nu\rho\sigma} = \partial_{(\mu} \xi_{\nu\rho\sigma)}\, ,
\ee
with {\it unconstrained}  infinitesimal rank-$3$ symmetric tensor parameter $\xi$.

We shall begin by confirming,  using the  ``canonical'' methods of \cite{Deser:2009hb,Andringa:2009yc},  that the equations
(\ref{sqrtFP4}) and (\ref{FPequiv4}) propagate, respectively, one and two massive modes. This analysis will be useful when we later turn to a similar analysis of the actions.   As we focus  on the  canonical structure of the equations, we make a time/space split for the components of the various fields, setting $\mu=(0,i)$ where $i=1,2$.  We may then choose a gauge such that\footnote{The summation convention applies.}
\be
\partial_i h_{i \mu\nu\rho} =0\, .
\ee
In this gauge we may write the components of $h$ in terms of five independent gauge-invariant fields
$(\varphi_0,\varphi_1,\varphi_2,\varphi_3,\varphi_4)$ as follows:
\begin{eqnarray}
h_{0000} &=&\frac{1}{\left( \nabla ^{2}\right) ^{2}}\varphi _{0}\text{ ,} \qquad
h_{000i} = \frac{1}{\left( \nabla ^{2}\right) ^{2}}\hat{\partial}
_{i}\varphi _{1}\text{ ,} \qquad
h_{00ij} = \frac{1}{\left( \nabla ^{2}\right)
^{2}}\hat{\partial}_{i}\hat{
\partial}_{j}\varphi _{2}\text{ ,}  \notag \\
h_{0ijk} &=&\frac{1}{\left( \nabla ^{2}\right)
^{2}}\hat{\partial}_{i}\hat{
\partial}_{j}\hat{\partial}_{k}\varphi _{3}\text{ ,}  \qquad
h_{ijkl} = \frac{1}{\left( \nabla ^{2}\right)
^{2}}\hat{\partial}_{i}\hat{
\partial}_{j}\hat{\partial}_{k}\hat{\partial}_{l}\varphi _{4}\text{ ,
}\label{dec}
\end{eqnarray}
where
\begin{equation}
\hat{\partial}_{i}=\varepsilon _{0}{}^{ij}\partial _{j}\text{ . }
\end{equation}
Note that we permit {\it space} non-locality, since this does not affect the canonical structure.
Substitution into (\ref{fieldstrength}) gives
 \begin{eqnarray}\label{g4}
G_{0000}(h) &=& (\nabla^2)^2 \varphi_4 \, , \qquad G_{000i}(h) =
 \nabla^2 \left(\hat\partial_i \varphi_3 + \partial_i \dot\varphi_4\right) \text{ ,} \nonumber \\
G_{00ij}(h) &=&    \left(\hat\partial_i\hat\partial_j \varphi_2 + 2 \hat\partial_{(i}\partial_{j)} \dot\varphi_3 + \partial_i\partial_j \ddot \varphi_4\right)\text{ ,} \nonumber \\
G_{0ijk}(h) &=& \frac{1}{\nabla^2} \left[ \hat\partial_i\hat\partial_j\hat\partial_k \varphi_1 + 3 \hat\partial_{(i} \hat\partial_j\partial_{k)}\dot\varphi_2 + 3 \hat\partial_{(i} \partial_j\partial_{k)} \ddot\varphi_3 +
\partial_i\partial_j\partial_k \left(\partial_t^3 \varphi_4\right) \right] \text{ ,} \nonumber \\
G_{ijkl}(h)&=&  \frac{1}{ (\nabla^2)^2}\left[ \hat\partial_i\hat\partial_j\hat\partial_k\hat\partial_l \varphi_0 + 4 \hat\partial_{(i}\hat\partial_j\hat\partial_k\partial_{l)} \dot\varphi_1 +  6\hat\partial_{(i}\hat\partial_j\partial_k\partial_{l)} \ddot\varphi_2  \right. \nonumber\\
&& \left. \qquad   \ +\,  4\hat\partial_{(i}\partial_j\partial_k\partial_{l)} \left(\partial_t^3 \varphi_3\right) +
 \partial_i\partial_j\partial_k\partial_l \left(\partial_t^4 \varphi_4\right)
\right] \text{ ,}
\end{eqnarray}
and hence
\begin{eqnarray}
G^{\text{tr}}_{00}(h) &=&  \nabla^2\left(\varphi_2 - \square\varphi_4\right)\text{ ,} \nonumber \\
G^{\text{tr}}_{0i}(h) &=& \hat\partial_i \left(\varphi_1 - \square\varphi_3\right) +
\partial_i \left(\dot\varphi_2 - \square \dot\varphi_4\right) \text{ ,} \\
G^{\text{tr}}_{ij}(h) &=& \frac{1}{\nabla^2} \left[ \hat\partial_i\hat\partial_j \left(\varphi_0 - \square\varphi_2\right) + 2 \hat\partial_{(i}\partial_{j)} \left(\dot\varphi_1 - \square\dot\varphi_3\right) + \partial_i\partial_j \left(\ddot\varphi_2 - \square \ddot\varphi_4\right)\right]
\text{ .} \nonumber
\end{eqnarray}

Using these results, the tensor equation $G^{\text{tr}}=0$ implies that
\be
\varphi_0 = \square^2\varphi_4\, , \qquad \varphi_1= \square \varphi_3\, , \qquad \varphi_2= \square\varphi_4\, ,
\ee
which eliminates $(\varphi_0,\varphi_1,\varphi_2)$ as {\it independent}  fields.  The dynamical equation of (\ref{sqrtFP4}) is then equivalent to
\be
\varphi_3 = \mu\varphi_4 \, , \qquad  \left(\square -\mu^2\right)\varphi_4 =0\, ,
\ee
so a single mode of mass $\mu$ is propagated. The dynamical equation of (\ref{FPequiv4}) is similarly equivalent to
\be\label{spin4red}
\left(\square -m^2\right)\varphi_3 =0\, , \qquad \left(\square -m^2\right)\varphi_4 =0\, ,
\ee
which shows that there are two propagating degrees of freedom of equal mass $m$.

The spins of the propagated modes cannot be determined easily by this method  since they are  defined with respect to the Lorentz transformations that leave invariant the original equations, and this has been broken by the gauge-fixing condition and subsequent
(space non-local) field redefinitions\footnote{In light of this, it is notable that the resulting equations  are still Lorentz invariant, but {\it these} Lorentz transformations are not those of the original action.}. However, the initial construction, which guarantees on-shell equivalence to (according to the case) the $\sqrt{\mathrm{FP}}$ or FP  equations for spin $4$,  tells us that the  modes have spin $4$.

\section{A Spin-$4$ analog of TMG}

We now seek a gauge-invariant, and manifestly Lorentz-invariant,  action that yields the `topologically-massive  spin-$4$ equations (\ref{sqrtFP4}). One can show that such an action must involve additional ``auxiliary''  fields that are set to zero by the equations of motion. There is a systematic procedure that can be used to find these auxiliary fields but here we just give the final result. One needs an auxiliary symmetric tensor $\pi_{\mu\nu}$ and an auxiliary vector field $\phi_\mu$ :
\begin{eqnarray} \label{spin4TMG}
S &=&\int d^3x\,\bigg\{\frac{1}{2\mu^{4}}h^{\mu\nu\rho\sigma}G_{\mu\nu\rho\sigma }\left(
h\right)  -\frac{1}{2\mu^{5}}h^{\mu\nu\rho\sigma}\varepsilon_{\mu}
{}^{\alpha\beta}\partial_{\alpha}G_{\beta\nu\rho\sigma}\left(  h\right)\nonumber \\
& & +\frac{1}{\mu^{4}}\pi^{\mu\nu}G_{\mu\nu}^{\text{tr}}\left(  h\right)
+\frac{1}{\mu^{3}}\pi^{\mu\nu}C_{\mu\nu}\left(  \pi\right)
+\frac{2}
{\mu^{2}}\pi^{\mu\nu}G_{\mu\nu}\left(  \pi\right) \nonumber \\
& & +\frac{4}{\mu}\pi^{\mu\nu}\varepsilon_{\mu}{}^{\alpha\beta}\partial_{\alpha}
\pi_{\beta\nu}+8\left(  \pi^{\mu\nu}\pi_{\mu\nu}-\pi^{2}\right) -\frac{1}{\mu}\phi^{\mu}\partial_{\mu}\pi+\frac{1}{\mu}\phi^{\mu}
\partial^{\alpha}\pi_{\alpha\mu}\nonumber \\
& &
-\frac{1}{16\mu}\phi^{\mu}\varepsilon_{\mu}{}^{\alpha\beta}\partial_{\alpha
}\phi_{\beta}+\frac{1}{4}\phi^{\mu}\phi_{\mu}\bigg\}\,,
\end{eqnarray}

\noindent where
\be
\pi = \eta^{\mu\nu}\pi_{\mu\nu}
\ee
and where $C_{\mu \nu}(\pi)$ denotes the Cotton tensor of $\pi$:
\be
C_{\mu \nu}(\pi) = \varepsilon_{(\mu}{}^{\alpha \beta} \partial_{|\alpha} S_{\beta| \nu)}(\pi) \,, \qquad S_{\mu \nu}(\pi) = G_{\mu \nu}(\pi) - \frac12 \eta_{\mu \nu} G^{\mathrm{tr}}(\pi) \,.
\ee
We now summarize how the equations of motion of this action may be shown to be equivalent to the equations (\ref{sqrtFP4}). We first write down the un-contracted equations of motion of $h_{\mu \nu \rho \sigma}$, $\pi_{\mu \nu}$ and $\phi_\mu$. These are equations with 4, 2 and 1 indices. We next construct out of these equations all possible equations with fewer indices by taking divergences and/or traces. In total, this leads to one (zero, two, two, four) equations with four (three, two, one, zero) indices. We now first use the four equations with zero indices to derive that
\be
\eta^{\mu \nu} G^{\mathrm{tr}}_{\mu \nu}(h) = \partial^\rho \partial^\sigma \pi_{\rho \sigma} = \pi = \partial^\mu \phi_\mu = 0 \,.
\ee
Next, we use the two equations with one index to show that $\partial^\lambda \pi_{\lambda \mu} = \phi_\mu = 0$. From the two equations with two indices we can then subsequently deduce that
\be
G^{\mathrm{tr}}_{\mu \nu}(h) = \pi_{\mu \nu} = 0 \,.
\ee
Substituting all these equations back into the original equation of motion for the tensor potential $h$ then leads to the equations (\ref{sqrtFP4}).

\section{A Spin-$4$ analog of NMG}\label{sec:NMG}

Similarly, we now seek a gauge-invariant, and manifestly Lorentz-invariant,  action that yields the spin-$4$ equations (\ref{FPequiv4}). In this case
the auxiliary vector field $\phi_\mu$ is not needed; only the symmetric tensor $\pi_{\mu\nu}$ and an
 additional scalar $\phi$ are required.  Defining
\be
G_{\mu\nu}(\pi) = \varepsilon_\mu{}^{\tau\rho} \varepsilon_{\nu}{}^{\eta\sigma} \partial_\tau\partial_\eta \, \pi_{\rho\sigma}\, ,
\ee
we may write the action as
\begin{eqnarray}\label{linearizeds=4NMG}
S &=&\int d^3x\,\bigg\{-\frac{1}{2m^{4}}h^{\mu \nu \rho \sigma }G_{\mu \nu \rho
\sigma }\left( h\right) +\frac{1}{2m^{6}}h^{\mu \nu \rho \sigma }\Box G_{\mu
\nu \rho \sigma }\left( h\right) \nonumber \\ [.15truecm]
&&+\frac{1}{m^{4}}\pi ^{\mu \nu }G^{\text{tr}}_{\mu \nu }\left( h\right)
-\frac{1}{2m^2}\pi^{\mu\nu}G_{\mu\nu}(\pi) - \frac{1}{2}\left(\pi^{\mu\nu}\pi_{\mu\nu}-\pi^2\right)\nonumber \\ [.15truecm]
&&+\phi \pi
+\frac{13}{12}\phi ^{2}+\frac{1}{12m^{2}}\phi \Box \phi \bigg\}\,.
\end{eqnarray}
Following the same procedure as in the TMG-like  case,  we summarize how the equations of motion of this action are equivalent to the equations (\ref{FPequiv4}).
We first write down the un-contracted equations of motion of $h_{\mu\nu\rho\sigma}\,, \pi_{\mu\nu}$ and $\phi$.
These are equations with 4, 2 and 0 indices. We next construct out of these equations all possible equations with fewer indices by
taking divergences and/or traces. In total, this leads to one (zero, two, one, four) equations with four (three, two, one, zero) indices.
The fact that there is no equation with three indices follows from the Bianchi identity
\be\label{Bianchi4}
\partial^\mu G_{\mu\nu\rho\sigma}(h) \equiv 0 \text{ ,}
\ee
which is the $s=4$ case of (\ref{Bianchi}). We now first use the four equations with zero indices to derive that
\begin{equation}
\eta^{\mu\nu} G^{\text{tr}}_{\mu\nu}(h) = \partial^\rho\partial^\sigma\pi_{\rho\sigma} = \pi = \phi =0\, .
\end{equation}
Next, we use the single equation with one index to show that $\partial^\lambda\pi_{\lambda\mu}=0$. From the two equations with
two indices we subsequently deduce that
\begin{equation}
G_{\mu\nu}^{\text{tr}}(h) = \pi_{\mu\nu}=0\,.
\end{equation}
Substituting all these equations back into the original equation of motion for the tensor potential $h$ then leads to
(\ref{FPequiv4}).


By construction, the action (\ref{linearizeds=4NMG}) propagates two spin-$4$ modes. We now need to show that both these modes are physical, rather than ghosts. To do this we first need to rewrite the action in terms of gauge-invariant variables only and then eliminate auxiliary fields to get an action for the propagating physical modes only. We have already seen how to write the gauge potential $h$ in terms of the five gauge-invariant fields $(\varphi_0,\varphi_1,\varphi_2,\varphi_3,\varphi_4)$. The auxiliary tensor $\pi_{\mu\nu}$ has six independent components, which we may write in terms of six independent fields $(\psi_0,\psi_1,\psi_2;\lambda_0,\lambda_1,\lambda_2)$ as follows:
\begin{eqnarray}
\pi _{00} &=&-\frac{1}{\nabla ^{2}}\left( \psi _{0}+2\dot{\lambda}
_{0}\right) \text{ ,}   \qquad
\pi _{0i} = -\frac{1}{\nabla ^{2}}\left[ \hat{\partial}_{i}\left(
\psi _{1}+ \dot{\lambda}_{1}\right) +\partial _{i}\left( \lambda
_{0}+\dot{\lambda}
_{2}\right) \right] \text{ ,}  \notag \\
\pi _{ij} &=&-\frac{1}{\nabla ^{2}}\left(
\hat{\partial}_{i}\hat{\partial} _{j}\psi
_{2}+2\hat{\partial}_{(i}\partial _{j)}\lambda _{1}+2\partial
_{i}\partial _{j}\lambda _{2}\right) \text{ .}
\end{eqnarray}
The dependence on the variables $(\lambda_1,\lambda_2,\lambda_3)$ is that of a  spin-$2$ gauge transformation, so the
tensor $G_{\mu\nu}(\pi)$, which is invariant under such a transformation, depends only on the three variables $(\psi_0,\psi_1,\psi_2)$.
Specifically, substituting the above expressions for the components of $\pi_{\mu\nu}$ gives
 \begin{eqnarray}
 G_{00}(\pi) &=& - \nabla^2 \psi_2\, , \qquad G_{0i}(\pi) = -\left(\hat\partial_i \psi_1 + \partial_i\dot\psi_2\right)\, , \nonumber \\
 G_{ij}(\pi) &=& -\frac{1}{\nabla^2} \left(\hat\partial_i \hat\partial_j \psi_0 + 2 \hat\partial_{(i}\partial_{j)} \dot\psi_1 + \partial_i\partial_j \ddot\psi_2\right)
 \end{eqnarray}
 and hence
 \be
\eta^{\mu\nu} G_{\mu\nu} (\pi) = -\left(\psi_0 - \square \psi_2\right)\, .
\ee

We are now in a position to determine the form of the action in terms of the gauge-invariant variables. Direct substitution yields the result
\be
S= \int d^3x \left\{ {\cal L}_1 + {\cal L}_2\right\}\, ,
\ee
where
\begin{eqnarray}
\mathcal{L}_{1} &=& \frac{4}{m^{6}}\varphi _{1}\left( \Box -m^{2}\right)
\varphi _{3}-\frac{2}{m^{4}}\psi _{1}\left( \varphi _{1}-\Box
\varphi _{3}\right) - \frac{1}{m^{2}}\psi _{1}^{2}-\,\lambda_{1}^{2} \nonumber \\
&& \quad  - \ \left( \psi _{1}+\dot{\lambda }_{1}\right) \frac{1}{\nabla
^{2}}\left( \psi _{1}+\dot{\lambda}_{1}\right) \text{ ,}
\label{Lagrangian part 1}
\end{eqnarray}
which depends only on the four fields $(\varphi_1,\varphi_3;\psi_1; \lambda_1)$,  and
\begin{eqnarray}
\mathcal{L}_{2} &=&\frac{1}{m^{6}}\varphi _{0}\left( \Box -m^{2}\right)
\varphi _{4}+\frac{3}{m^{6}}\varphi _{2}\left( \Box -m^{2}\right) \varphi
_{2}  \notag \\
&&-\frac{1}{m^{4}}\,\psi _{0}\left( \varphi _{2}-\Box \varphi
_{4}\right) - \frac{1}{m^{4}}\,\psi _{2}\left( \varphi _{0}-\Box
\varphi _{2}\right) -
\frac{1}{m^{2}}\,\psi _{0}\psi _{2}  \notag \\
&&+2\lambda _{2}\psi _{2}-\left( \psi _{2}+2\lambda _{2}\right)
\frac{1}{
\nabla ^{2}}\left( \psi _{0}+2\dot{\lambda}_{0}\right)  \notag \\
&&-\,\left( \lambda _{0}+\dot{\lambda}_{2}\right) \frac{1}{\nabla
^{2}} \left( \lambda _{0}+\dot{\lambda}_{2}\right) +\phi
\frac{1}{\nabla ^{2}}
\left( \psi _{0}+2\dot{\lambda}_{0}\right)  \notag \\
&&-\,\phi \psi _{2}-2\phi \lambda _{2}+\frac{13}{12}\phi
^{2}+\frac{1}{ 12m^{2}}\phi \Box \phi \text{ , }  \label{Lagrangian
part 2}
\end{eqnarray}
which depends only on the remaining eight fields $(\varphi_0,\varphi_2,\varphi_4; \psi_0,\psi_2; \lambda_0,\lambda_2; \phi)$.
We have already seen that the propagating fields are $\varphi_3$ and $\varphi_4$, so it must be that one spin-$4$ mode is propagated by each of these two parts of the action. We now aim to confirm this and to determine whether the propagated modes are physical or ghosts. A systematic analysis is possible but we give only the final results.

Discarding total derivatives, the Lagrangian ${\cal L}_1$ can be rewritten as
\begin{equation}\label{L1}
\mathcal{L}_{1}=\frac{1}{m^{8}}\tilde{\varphi}_{1}\left( \Box -m^{2}\right)\tilde{\varphi}_{1}
+\tilde{\psi}_{1}^{2}-\tilde{\lambda}_{1}^{2}+\frac{1}{
m^{2}}\tilde{\varphi}_{3}^{2}\text{ ,}
\end{equation}
where
\begin{eqnarray}\label{redefs}
\tilde \varphi_1 &=&  \varphi _{1} + m^2\varphi_3
+\frac{1}{2}m^{2}\psi _{1}  \text{ ,}\qquad
\tilde\varphi_3 = \varphi_3 - \frac{1}{m^2} \varphi_1 - \frac{1}{2} \psi_1 \text{ ,} \nonumber \\
\tilde{\psi}_{1} &=&  \frac{1}{\sqrt{-\nabla ^{2}}}\left[ \left(1+ \frac{\nabla^2}{2m^2} \right)\psi
_{1}+\dot{ \lambda}_{1}+\frac{\nabla
^{2}}{m^{4}}\varphi_{1}-\frac{\nabla ^{2}}{ m^{2}}\varphi
_{3}\right] \text{ ,} \nonumber \\
\tilde{\lambda}_{1} &=& \lambda _{1}+\frac{1}{m^{4}}\dot{\varphi}_{1} + \frac{1}{2m^2} \dot\psi_1 -\frac{1}{m^{2}}\dot{\varphi}_{3}\text{ .}
\end{eqnarray}
Using these relations, the field equations of (\ref{L1}) can be shown to imply that $\psi_1=\lambda_1=0$ and
\be
\varphi_1 = \square\varphi_3\, , \qquad \left(\square -m^2\right)\varphi_3 =0\, ,
\ee
in agreement with our earlier conclusion that $\varphi_3$ is the only independent propagating field (in the original basis).

In a similar way, the Lagrangian ${\cal L}_2$ can be rewritten as
\begin{equation}
\mathcal{L}_{2}=\frac{4}{m^{6}}\tilde\varphi_{2}\left( \Box
-m^{2}\right)
\tilde\varphi_{2}-\tilde{\phi}\, \tilde{\varphi}_{4}-\frac{1}{m^{4}}\tilde{\psi}
_{2}\, \tilde{\varphi}_{0}+\tilde{\lambda}_{2}\tilde{\psi}_{0}+\tilde{\lambda}
_{0}^{2}\text{ ,}
\end{equation}
where
\begin{eqnarray}
\tilde\varphi_0 &=&  \varphi_0 -m^2 \varphi_2 + m^2\left(\square + m^2\right)\psi_2 - m^2\square\phi - \frac{7}{6} m^4 \phi
 \text{ ,}  \qquad
\tilde\varphi_{2} = \varphi _{2}+\frac{1}{6}m^{2}\phi \text{ ,} \nonumber \\
\tilde{\varphi}_{4} &= & \varphi_{4}-\frac{1}{m^{2}}\varphi_{2} - \frac{1}{6}\phi\text{ ,} \qquad
\tilde{\psi}_{0} = -\frac{1}{\nabla ^{2}}\left( \psi _{0}-\Box \psi
_{2}+\Box \phi \right) \text{ ,} \nonumber \\
\tilde{\psi}_{2} &=& \psi _{2}-\frac{1}{m^{2}}\left( \Box -m^{2}\right)
\varphi _{4}\text{ ,} \qquad
\tilde{\lambda}_{0} = \frac{1}{\sqrt{-\nabla ^{2}}}\left(
\lambda _{0}- \dot{\psi}_{2}-\dot{\lambda}_{2}+\dot{\phi}\right)
\text{ ,} \nonumber \\
\tilde{\lambda}_{2} &=&2\lambda _{2}+\frac{\nabla ^{2}}{m^{4}}\left(
\varphi _{2}-\Box \varphi _{4}\right) +\left(1 + \frac{\nabla^2}{m^2}\right) \psi_{2}-\phi \text{ ,} \nonumber \\
\tilde{\phi} &=&\frac{7}{6} \phi - \frac{1}{6m^2} \square\phi - \frac{1}{m^4} \left(\square -m^2\right) \varphi_2 - \psi_2
\text{ .}
\end{eqnarray}
Using these relations the field equations of ${\cal L}_2$ can be shown to be equivalent to $\psi_0=\psi_2=\lambda_0=\lambda_2=\phi=0$ and
\be
\varphi_0= \square^2\varphi_4\, , \qquad \varphi_2 = \square \varphi_4\, , \qquad
\left(\square-m^2\right)\varphi_4=0 \text{ ,}
\ee
again in agreement with our earlier conclusion that $\varphi_4$ is the only independent propagating field (in the original basis).

If we now recombine the two Lagrangians ${\cal L}_1$ and ${\cal  L}_2$ and eliminate auxiliary fields we arrive at the Lagrangian
\begin{eqnarray}
\mathcal{L} = \frac{1}{m^{8}}\tilde{\varphi}_{1}\left( \Box -m^{2}\right)
\tilde{\varphi
}_{1}\mathcal{+}\frac{4}{m^{6}}\tilde{\varphi}_{2}\left( \Box
-m^{2}\right) \tilde{\varphi}_{2}
\text{ .}
\end{eqnarray}
Observe that both terms have the same sign. This means that the overall sign can be chosen such that both modes are physical. In our conventions, the sign that we have chosen is precisely such that this is the case, so our spin-$4$ action is ghost-free.

\section{Conformal spin $4$}

So far, we have considered massive spin-4 gauge theories with equations that can be obtained by solving the differential subsidiary condition of corresponding FP or $\sqrt{\mathrm{FP}}$ equations. As mentioned in section \ref{sec:intro}, it is possible to solve, simultaneously,  both the differential subsidiary constraint and the algebraic trace-free condition on the FP field, which thereby becomes a Cotton-type tensor for a gauge potential $h$ that is subject to a conformal-type linearized gauge transformation that  can be used to remove its trace. Here we present a few further details of this construction for spin $4$ and we analyze the physical content of the `conformal' models that one finds this way. 

The spin-$4$ FP field becomes the spin-$4$ Cotton-type tensor 
\be\label{StoC} 
C_{\mu \nu \rho \sigma} =
\varepsilon_{(\mu}{}^{\alpha_1 \beta_1}
\varepsilon_{\nu}{}^{\alpha_2 \beta_2}
\varepsilon_{\rho}{}^{\alpha_3 \beta_3} \partial_{|\alpha_1}
\partial_{\alpha_2} \partial_{\alpha_3} S_{\beta_1 \beta_2
\beta_3|\sigma)}(h)  \,, 
\ee 
where the spin-4 Schouten-type tensor $S$, for symmetric rank-$4$ tensor potential $h$, is 
\be S_{\mu \nu \rho
\sigma} (h)= G_{\mu \nu \rho \sigma}(h) - \eta_{(\mu \nu}
G^{\mathrm{tr}}_{\rho \sigma)}(h) + \frac18 \eta_{(\mu \nu}
\eta_{\rho \sigma)} \eta^{\alpha \beta} G^{\mathrm{tr}}_{\alpha \beta}(h) \,. 
\ee 
The conformal-type transformation for spin-$4$ is 
\be
\delta h_{\mu\nu\rho\sigma} =
\eta_{(\mu\nu}\, \Lambda_{\rho\sigma)}\,. \label{lctr} 
\ee
The  invariance of the Cotton-type tensor under this gauge transformation is an immediate consequence of the following 
simple transformation law for the spin-4 Schouten-type  tensor:
\be \delta S_{\mu \nu \rho \sigma}(h) = \partial_{(\mu} \partial_\nu \Omega_{\rho \sigma)} \,, \label{transfschouten}
\qquad \Omega_{\mu \nu} = G_{\mu \nu}(\Lambda) - \frac16 \eta_{\mu \nu} G^{\mathrm{tr}}(\Lambda) \, ,  
\ee
where the tensor $G(\Lambda)$ is the linearized Einstein tensor for the 2nd-rank tensor parameter $\Lambda$. 

Following the procedure outlined above, we deduce that the  spin-$4$ FP equations are equivalent to the single equation
 \be \label{conf4}
 (\Box - m^2) C_{\mu \nu \rho \sigma} = 0
\,, \ee 
with no lower-derivative constraints; what were the differential and algebraic constraints of the FP theory are now the Bianchi and trace-free {\it identities}
\be\label{confidentities}
\partial^\mu C_{\mu\nu\rho\sigma} \equiv 0\, , \qquad \eta^{\mu\nu} C_{\mu\nu\rho\sigma} \equiv 0\, .
\ee
The equation (\ref{conf4})  may integrated to the following 9th-order action, without the need of auxiliary fields:
\be S^{(9)} = \frac12 \int\, d^3 x \ h^{\mu \nu \rho
\sigma} (\Box - m^2) C_{\mu \nu \rho \sigma}(h) \,.\label{s=4ca} 
\ee
Similarly,  the  $\sqrt{\mathrm{FP}}$ spin-$4$ equations become
\be
\varepsilon_{\mu}{}^{\alpha \beta} \partial_\alpha C_{\beta \nu \rho \sigma} = \mu C_{\mu \nu \rho \sigma} \,,
\ee
again with no lower-derivative constraints. These equations can be integrated to the following action, 8th order in derivatives:
\be
S^{(8)} = \frac12 \int\, d^3 x \ \left\{ h^{\mu \nu \rho \sigma} C_{\mu \nu \rho \sigma}(h) + \frac{1}{\mu} \varepsilon_\mu{}^{\alpha \beta} h^{\mu \nu \rho \sigma} \partial_\alpha C_{\beta \nu \rho \sigma}(h) \right\} \,.
\ee
Again, no auxiliary fields are needed. 

By construction, the actions $S^{(8)}$ and $S^{(9)}$ propagate, respectively, one or two modes of spin $4$, although there is 
no guarantee that none of the modes is a ghost. To settle this issue we may use the additional  gauge invariance (\ref{lctr}) to go to a 
gauge in which 
\be h_{ii\mu\nu}=0\,. \ee 
In this gauge the only non-zero components appearing in the  decomposition \eqref{dec} are
\be
h_{0000} = \frac{1}{(\nabla^2)^2} \, \varphi_0 \, , \qquad h_{000i} = \frac{1}{(\nabla^2)^2} \, \hat\partial_i \varphi_1\, . 
\ee
As a consequence of this simplification, we will need only the following components of  the Cotton-type tensor
\begin{eqnarray} C_{0000} &=& -\frac12 (\nabla^2)^2 \varphi_1 \nonumber 
\,,\\[.25truecm]
C_{000i} &=&  - \frac18 \nabla^2 \hat{\partial}_i \varphi_0 -\frac12 \nabla^2 \partial_i \dot{\varphi}_1 \,,\nonumber \\
C_{00ij} &=& - \frac{1}{2} \hat\partial_i\hat\partial_j \Box \varphi_1 - \frac{1}{4} \hat\partial_{(i}\partial_{j)} \dot\varphi_0
- \frac{1}{2} \partial_i\partial_j \ddot \varphi_1\, . 
\end{eqnarray}
The remaining components are not zero but they are determined in terms of the ones given by the conditions (\ref{confidentities}). 

Using these results for the Cotton-type tensor we find that 
\be S^{(9)} \ = \ -\frac12 \int\! d^3x\
\varphi_0\big(\Box-m^2\big)\varphi_1 \, .\ee
This action propagates two modes but one is a ghost. This was to be expected because this is what happens for
$s=2,3$ \cite{Bergshoeff:2009tb}.  In contrast $S^{(8)}$ propagates a single mode. To check this we may use 
the above expressions for the Cotton-type tensor components to deduce that 
\be
S^{(8)} = \ \int\! d^3x\, \left\{  -\frac12 \varphi_0 \varphi_1 + \frac{1}{16\mu} \varphi_0^2 + \frac{1}{\mu} \varphi_1 \Box \varphi_1\right\}
\ee
The field $\varphi_0$ is now auxiliary and may be trivially eliminated, so that 
\be
S^{(8)} \quad \rightarrow \ \frac{1}{\mu} \int\! d^3\,x\ \varphi_1 \left(\Box -\mu^2\right) \varphi_1\, . 
\ee
This  action clearly propagates a single mode, and this mode is physical if $\mu>0$. We have therefore found an $8$th order ghost-free action that propagates a single spin-$4$ mode.

\section{Conclusions}

In this paper, we have constructed ghost-free actions that yield spin-4 analogs of linearized massive gravity models.  One, of 5th order in derivatives, is a parity-violating field theory that propagates a single spin-$4$ mode; it is a spin-$4$ analog of  linearized ``topologically massive gravity'' (TMG). The other, of 6th order in derivatives, is a parity-preserving  field theory that propagates two spin-$4$ modes; it is a spin-$4$ analog of ``new massive gravity'' (NMG). In both cases the action involves auxiliary fields and is invariant under an {\it unconstrained} spin-$4$ gauge transformation (i.e. one in which the 3rd-rank symmetric tensor gauge parameter is not constrained to be trace-free). The absence of ghosts is non-trivial in the NMG-type case but we have verified by ``canonical'' methods that both propagating modes are physical.

In the spin-$2$ case, both TMG and NMG are particular limits of  a ``general massive gravity'' model that propagates two spin-$2$ modes, generically with different masses \cite{Bergshoeff:2009hq}.  We expect there to exist a spin-$4$ analog of this model, such that the TMG-type and NMG-type spin-$4$ models constructed here arise as special cases.

Given a spin-$s$ TMG-type model, we could construct a parity preserving theory by taking the action to be the sum of two TMG-type models with opposite sign masses. This  bi-field model has the same propagating content as a single NMG-type model but is one order lower in derivatives.  In the spin-$2$ case there exists a ``soldering'' procedure that allows one to convert the bi-field TMG model into an NMG model \cite{Dalmazi:2010bf}.  We do not expect this to work for spin $3$ (because the attempt to construct a spin-$3$ NMG-type model, along the lines of this paper, yields a model with ghosts \cite{Bergshoeff:2009tb})  but there might exist some analogous ``soldering'' procedure for spin $4$.

We have also constructed a parity-violating ghost-free ``conformal spin-$4$'' action that propagates a single spin-$4$ mode. In this case the action  is 8th order in derivatives but invariant under a spin-$4$ analog of a spin-$2$ linearized conformal gauge invariance, in addition to the unconstrained spin-$4$ gauge invariance. This is the spin-$4$ analog of  ``new topologically massive gravity''.  There is a parity-preserving version, of 9th-order in derivatives, that propagates two spin-$4$ modes but one mode is a ghost.  However a parity-preserving bi-field model of 8th order  will have the same physical content as the 6th-order spin-$4$ NMG-type action.

Of course, what is ultimately of importance is which, if any,  of the various models constructed here has some
extension to an interacting theory or, more likely, plays a role in the context of some interacting 3D  theory of higher
spins.  It seems likely to us that a much improved understanding of the general spin $s>2$ case will be needed to begin addressing this issue. This  lies outside the scope of the present paper.  However, we will conclude with an argument that goes some way towards a proof of the conjecture in \cite{Bergshoeff:2009tb} that an NMG-like action for integer spin $s$ is ghost-free only if $s$ is even.

To prove the conjecture, we should start from an action for a spin-$s$ NMG-type model, as found here for $s=4$, in which case we would  first have to find the auxiliary fields. Recall that these auxiliary fields are needed to impose the lower-order constraint equation. As a shortcut, we could construct an action for the dynamical equation alone, for which auxiliary fields are not needed, and then impose `by hand' the constraint equation. In other words, we consider the Lagrangian
\begin{equation}\label{truncated-spin-s}
\mathcal{L}_{\text{spin-}s}=\frac{1}{2}h^{\mu _{1}\cdots \mu _{s}}\left(
\Box -m^{2}\right) G_{\mu _{1}\cdots \mu _{s}}\left( h\right) \text{ .}
\end{equation}
To the field equations we must now add, `by hand',  the trace-free constraint
\begin{equation}
G_{\mu _{1}\cdots \mu _{s-2}}^{\text{tr}}\left( h\right) =0\,.
\label{spin-general traceless condition}
\end{equation}
We now proceed to a canonical analysis of this Lagrangian, and the constraint, by setting
\begin{equation}
h_{i_{1}i_{2}\cdots i_{t}0\cdots 0} = \frac{1}{\left( -\nabla ^{2}\right)
^{s/2}}\hat{\partial}_{i_{1}}\cdots \hat{\partial}_{i_{t}}\varphi _{t}\text{
,}  \qquad (t=0,\dots,s).
\end{equation}
It then follows, for $r=0,\cdots ,s$,  that
\begin{equation}
G_{i_{1}\cdots i_{r}0\cdots 0}\left( h\right) =\left( -1\right) ^{r}\left(
-\nabla ^{2}\right) ^{\frac{s}{2}-r}\sum_{p=0}^{r}\dbinom{r}{p}\hat{\partial}
_{(i_{1}}\cdots \hat{\partial}_{i_{p}}\partial _{i_{p+1}}\cdots \partial
_{i_{r})}\partial _{0}^{r-p}\varphi _{s-p}\, ,
\end{equation}
and, for $r =2,\cdots ,s$, that
\begin{eqnarray}
G_{i_{1}\cdots i_{r-2}0\cdots 0}^{\text{tr}}\left( h\right) &=& \left(
-1\right) ^{r+1}\left( -\nabla ^{2}\right) ^{\frac{s}{2}-r+1} \times  \\
&& \sum_{p=0}^{r-2}
\dbinom{r-2}{p}\hat{\partial}_{(i_{1}}\cdots \hat{\partial}_{i_{p}}\partial
_{i_{p+1}}\cdots \partial _{i_{r-2})}\partial _{0}^{r-p-2}\left( \varphi
_{s-p-2}-\Box \varphi _{s-p}\right)\, .  \nonumber
\end{eqnarray}
Substituting  into the Lagrangian (\ref{truncated-spin-s}) we obtain
\begin{equation}\label{lags}
\mathcal{L}_{\text{spin-}s} = \left\{
\begin{array}{ccc}
\frac{1}{2} \dbinom{s}{s/2} \varphi _{s/2}\left(
\Box -m^{2}\right) \varphi _{s/2}+\sum_{t=0}^{\left(s/2\right) -1}\dbinom{s
}{t}\varphi _{t}\left( \Box -m^{2}\right) \varphi _{s-t} && {\rm even} \ s \, ,
\\
-\sum_{t=0}^{\left( s-1\right) /2}
\dbinom{s}{t}\varphi _{t}\left( \Box -m^{2}\right) \varphi _{s-t} && {\rm odd} \ s \, .
\end{array}\right.
\end{equation}
In either case, the equations of motion that follow from this Lagrangian are
\begin{equation}
\left( \Box -m^{2}\right) \varphi _t =0\, ,  \qquad (t=0,\cdots ,s)\, .
\label{spin-general e.o.m. canonical}
\end{equation}
To these equations we have to add the trace-free condition (\ref{spin-general traceless condition}), which is equivalent to
\begin{equation}
\varphi _{s-p-2}=\Box \varphi _{s-p}\, , \qquad (p=0,\cdots ,s-2)\, .
\label{spin-general traceless condition canonical}
\end{equation}
By combining (\ref{spin-general e.o.m. canonical}) with (\ref{spin-general
traceless condition canonical}) one finds, for  $t\leq s$, that
\begin{eqnarray}
\varphi _0 &=& m^t\varphi _t \, ,  \ \ \ \, \qquad  t=0,2,4,\cdots  \, ,\\
 \varphi _1 &=& m^{t-1}\varphi _t \, , \, \qquad  t=1,3,5,\cdots \, .
\end{eqnarray}

Next, one substitutes these equations into the Lagrangians of (\ref{lags}) in order to
eliminate all fields other than $\varphi _{0}$ and $\varphi _{1}$. For even spin $s$ the resulting Lagrangians
contain only two terms:  $\varphi _{0}\left( \Box -m^{2}\right) \varphi _{0}$
and $\varphi _{1}\left( \Box -m^{2}\right) \varphi _{1}$, both with the same (positive)
sign. The even spin Lagrangians are therefore ghost-free. In contrast, the Lagrangians for
odd spin contain  only one off-diagonal term,  which is proportional to $\varphi _{0}\left( \Box
-m^{2}\right) \varphi _{1}$. In this case, therefore, one mode is physical and the other a ghost.
Although this argument  falls short of a proof that there is a ghost-free spin-$s$ NMG-type action
only for even $s$,  we believe that it captures the essential difference between the even and odd spin cases.

\section*{Acknowledgements} This research was supported in part by the Perimeter Institute for Theoretical Physics. PKT thanks the Perimeter Institute for hospitality during the completion of this paper. The work of JR is supported by the Stichting Fundamenteel Onderzoek der Materie (FOM). The work of YY is supported by the Ubbo Emmius Programme administered by the Graduate School of Science, University of Groningen.


\providecommand{\href}[2]{#2}\begingroup\raggedright\endgroup

\end{document}